\documentclass[aps, prl, twocolumn, nofootinbib, superscriptaddress]{revtex4-1}
\usepackage{amsmath}
\usepackage{amssymb}
\usepackage{graphicx}
\usepackage{latexsym}
\usepackage{amsfonts}
\usepackage{layout}
\usepackage{verbatim}
\usepackage{dcolumn}    
\usepackage{bm}         
\usepackage[T1]{fontenc}
\usepackage{bbold} 
\usepackage{mathrsfs}
\usepackage{mathtools}
\usepackage{graphicx}
\usepackage{tikz}
\usepackage{lipsum}
\usepackage{pgf}
\usepackage{xcolor}
\usepackage{hyperref}
\usepackage{tabularx}
\usepackage{cancel}
\hypersetup{colorlinks=true,linkcolor={red!50!black},citecolor={blue!50!black},urlcolor={blue!80!black}}
\usepackage[caption=false]{subfig}
\usepackage{braket}

\makeatletter
\newcommand{\cz}{
  \mathord{\mathpalette\vaggelis@z{z}}%
}
\newcommand{\cZ}{
  \mathord{\mathpalette\vaggelis@z{Z}}%
}
\newcommand{\vaggelis@z}[2]{%
  \sbox\z@{$\m@th#1#2$}%
  \ooalign{%
    $\m@th#1#2$\cr
    \hidewidth
    \vrule height \dimexpr.5\ht\z@+0.03ex\relax
           depth -\dimexpr.5\ht\z@-0.03ex\relax
           width .5\wd\z@
    \hidewidth\cr
  }%
  \vphantom{\box\z@}
}
\makeatother
\definecolor{myteal}{RGB}{31, 119, 180}   
\definecolor{myorange}{RGB}{255, 127, 14} 

\begin{document}
\title{Micromagnet-free operation of electron spin qubits in Si/Si$_{1-x}$Ge$_x$ vertical double quantum dots}

\author{Abhikbrata Sarkar}\affiliation{Department of Physics, University of Basel, Klingelbergstrasse 82, 4056 Basel, Switzerland}

\author{Daniel Loss} \affiliation{Department of Physics, University of Basel, Klingelbergstrasse 82, 4056 Basel, Switzerland}
\affiliation{Physics Department, King Fahd University of Petroleum and Minerals, 31261, Dhahran, Saudi Arabia}
\affiliation{Center for Advanced Quantum Computing, KFUPM, Dhahran, Saudi Arabia}
\affiliation{RDIA Chair in Quantum Computing}
\date{\today}

\begin{abstract} 
We study a vertical double quantum dot (DQD) in a Si/Si$_{1-x}$Ge$_x$/Si double-well heterostructure for full electrical control of electron Loss-DiVincenzo (LD) spin qubits, using realistic device modeling and numerical simulations. Due to the emerging spin-orbit interaction in the DQD, as well as strain from the gate electrodes, small (percentage range) but finite $g$-tensor variations emerge. In addition, we find a large valley splitting, on the order of $E_v{\sim}250 \,\mu$eV. As a result, multiple avenues for fast electrical single qubit rotations emerge. An ac electric field gives rise to electric dipole spin resonance (EDSR), while electron spin resonance (ESR) in the presence of an ac magnetic field can be electrically controlled by local gates due to varying $g$ factors in DQDs. We also show that  shuttling between neighboring dots, in vertical and horizontal direction, results in ultrafast single qubit gates of less than a nanosecond. Remarkably, this DQD architecture completely eliminates the need for micromagnets, significantly facilitating the scalability of LD spin qubits in semiconductor foundries.
\end{abstract}
\maketitle

{\it Introduction.}
Spin qubits in silicon are deemed highly coherent and scalable~\cite{loss1998quantum,levy2002universal,petta2005coherent,tyryshkin2012electron,burkard2023semiconductor} 
as well as the most integrable with existing semiconductor foundry processes~\cite{zwanenburg2013silicon,zhang2019semiconductor}.
Additional advantages of Si spin qubits include possible hyperfine-free isotopes~\cite{itoh2014isotope} and the absence of piezoelectric coupling to phonons~\cite{prada2008singlet}. Several Si architectures have been successfully implemented in the few-dot regime~\cite{chanrion2020charge,veldhorst2015two,veldhorst2017silicon,gilbert2020single,harvey2019spin,hollenberg2006two,he2019two,krauth2022flopping,watson2017atomically,Maurand2016,Liles2018,Piot2022,camenzind2022hole,yu2023strong}. Among these, Si/Si$_{1-x}$Ge$_x$  heterostructures are particularly attractive due to the achievement of a very high-mobility 2D electron gas (2DEG) in almost nuclear spin-free $^{28}$Si hosts~\cite{mi2015magnetotransport,melnikov2015ultra,huang2012mobility}. Epitaxially grown strained silicon quantum wells in Si/Si$_{1-x}$Ge$_x$ heterostructures offer a suitable platform for scalable quantum computation with electron spins~\cite{kim2014quantum,kawakami2014electrical,huang2017electrically}. These systems have been shown to host coherent electron spin qubits featuring high fidelity one- and two-qubit gates~\cite{kawakami2014electrical,kawakami2016gate,sigillito2019site,mills2022two,connors2022charge,yoneda2018quantum,xue2022quantum,philips2022universal,li2018crossbar,shi2011tunable,zajac2015reconfigurable}.

However, there are two major challenges. First, for electron LD spin qubits in silicon, micromagnets are required to perform single-qubit gates~\cite{kawakami2016gate,takeda2016fault,song2024coherence} because the intrinsic spin-orbit interaction (SOI) in pure Si is weak. This creates several issues, such as magnetic noise, cross-talk, and scalability~\cite{kha2015micromagnets,dumoulin2021low,undseth2023nonlinear,unseld2025baseband}. It would thus be highly desirable to eliminate the need for micromagnets in LD qubit architectures. Second, the conduction band of bulk silicon has six degenerate minima close to the X-point at $|\mathbf{k_0}|=0.85\,\frac{2\pi}{a_{\text{Si}}}$, known as valleys. For a heterostructure grown with Si/Si$_{1-x}$Ge$_x$, the tensile strain ($\varepsilon_{xx}{=}\varepsilon_{yy}{\approx}\,0.42 \,x\%$) in the silicon layer partially lifts the six-fold valley degeneracy by inducing an energy gap between the $\pm z$ and $\pm x, \pm y$ valleys. However, the valley splitting between the lowest lying $\pm z$ states is highly device-dependent, spanning from negligibly small values to several hundred $\mu$eV, reflecting a strong sensitivity to local electrostatics and interface disorder~\cite{borselli2011measurement,mi2017high,neyens2018critical}. Extensive theoretical modeling has been conducted to study the multivalley physics of Si conduction bands~\cite{Vrijen2000,Friesen2007,Gamble2015,Burkard2016,
Ferdous2018,klimeck2007atomisticI,klimeck2007atomisticII,abadillo2018signatures,dodson2022valley,saraiva2011intervalley,sarkar2022optimisation}, producing useful systematic strategies to lift valley degeneracy in practice~\cite{hosseinkhani2020electromagnetic,tariq2019effects,goswami2007controllable,culcer2010interface,boross2016control,adelsberger2024valley}. In experiments, several valley-splitting enhancement strategies have been implemented, e.g., interface engineering~\cite{paquelet2022atomic,Lodari2022,degli2024low,marcks2025valley}, periodic Ge concentration in wiggle-well~\cite{mcjunkin2022sige,feng2022enhanced,gradwohlenhanced2025}, or a single Ge `spike' in the quantum well~\cite{mcjunkin2021valley}.

Here, we theoretically demonstrate that we can simultaneously circumvent both problems: micromagnets and valley degeneracy, by utilizing vertical DQDs~\cite{burkard2000spin}  in a Si$_{0.7}$Ge$_{0.3}$/Si/Si$_{1-x}$Ge$_x$/Si/Si$_{0.7}$Ge$_{0.3}$ double quantum well structure hosting a single electron spin. 
First, using effective mass theory, we show that the vertical confinement, consisting of the two narrow Si wells separated by a thin Si$_{1-x}$Ge$_x$ layer, produces a large valley splitting. Realistic device simulation reveals that the DQD system showcases the tunability of the electron $g$ through the top (plunger) gate: $g=2\pm\mathcal{O}(10^{-2})$. In the presence of an ac electric field, the device dependent SOI can enable fast EDSR with a Rabi frequency of $\sim$ 20 MHz, without any micromagnet. For ESR with an ac magnetic field, the $g$ tensor tunability provides a powerful electrical control knob, as it allows for shifting the resonance frequency without changing the microwave field itself. Finally, we consider spin shuttling in vertical and horizontal DQDs and show that both can generate ultrafast single qubit rotations in subnanoseconds, owing to the angular difference in the principal axes of $g$ tensors. While vertical DQDs  have been experimentally realized for holes in Ge/SiGe double quantum wells \cite{tidjani2023vertical,ivlev2024coupled}, we are not aware of analogous structures for electrons proposed here.

{\it DQD Hamiltonian and device model.}
A single electron spin in the Si/Si$_{1-x}$Ge$_x$/Si DQD in a heterostructure grown in [001] can be described by a spin-valley Hamiltonian given in the basis $\{\left|+z, \uparrow\right\rangle, \left|+z, \downarrow\right\rangle, \left|-z, \uparrow\right\rangle, \left|-z, \downarrow\right\rangle\}$~\cite{woods2024g}. Its matrix representation is thus written in the form $\sigma_i\otimes\tau_j$, where $\sigma_i$ and $\tau_j$ ($i,j=\{0,x,y,z\}$) denote the Pauli matrices in the spin and valley energy subspaces, respectively,
\begin{equation}\label{eq1:spinvalleyHamiltonian}
    H_{\text{tot}}^{4\times 4}= H_{\text{EFA}}+ H_{\sigma\tau}+ eF_zz \,\mathbb{I}_{4\times 4}+\frac{\mu_B}{2}\,\mathbf{\sigma}\cdot\mathbf{g}\cdot\mathbf{B}\,\mathbb{I}_{2\times 2}.
\end{equation}
The first term in Eq.~\ref{eq1:spinvalleyHamiltonian} denotes the effective mass Hamiltonian describing the electron dispersion in the $\pm z$ valleys of the lowest conduction band of silicon,
\begin{eqnarray}\label{eq2:EFAHamiltonian}
    &&H_{\text{EFA}}=\biggl[\frac{\hbar^2}{2}\biggl(\frac{(k_z+eA_z(x,y)/\hbar)^2}{m_l}+\frac{k_x^2+k_y^2}{m_t}\biggr)\nonumber\\
    &&+ V(\mathbf{r})+ \Xi_u\varepsilon_{zz}\biggr]\,\mathbb{I}_{4\times 4}-2\Xi_u'\varepsilon_{xy}\,\mathbb{I}_{2\times 2}\,\tau_z.
\end{eqnarray}
Here, $m_l{=}0.916\,\times 9.11\times10^{-31}$ kg and $m_t{=}0.191\,\times 9.11\times10^{-31}$ kg denote the effective longitudinal and transverse masses of the $z$-valleys. Note that the shear strain $\varepsilon_{xy}$ induces a valley splitting term $\tau_z$ between the $+z$ and $-z$ valleys. For in-plane magnetic field operation, the vector potential is given by $A_z(x,y){=}B_xy{-}B_yx$. The confinement potential is given by $V(\mathbf{r}){=}V(x,y){+}V_{\text{DW}}(z)$, where $V(x,y)$ is modeled as 2D parabolic potential in $x{-}y$ directions, with characteristic lengths $L_x{=}L_y{=}L_{ip}$. The $z$-double well potential $V_{\text{DW}}(z)$ is modeled as a numerical function of the single Si $z$-well width $L$, the Si$_{1-x}$Ge$_x$ barrier width $a$, and the barrier height $V_b$ (see Supplementary Material (SM) \cite{suppl}).
\begin{figure}
    \centering
    \includegraphics[width=0.48\textwidth]{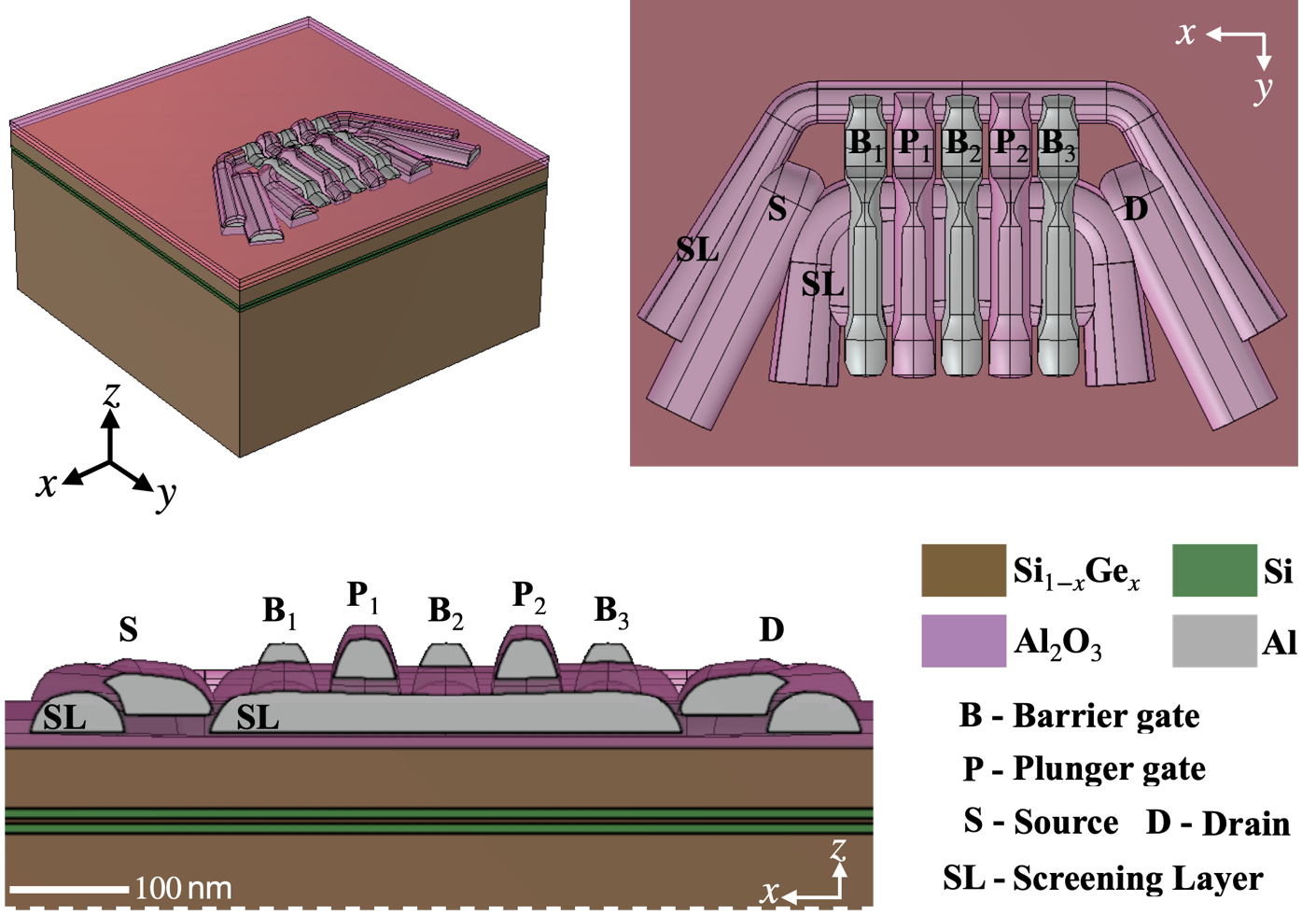}
    \vspace{-0.6 cm}
    \caption{{\bf Realistic device simulation outline.} Geometry of the simulated heterostructure with 300 nm regrowth Si$_{0.7}$Ge$_{0.3}$ layer, followed by the Si/Si$_{1-x}$Ge$_x$/Si double well structure, then 30 nm Si$_{0.7}$Ge$_{0.3}$ buffer layer. The tri-layer Al gates (from bottom: layer 1$\rightarrow$initial SL${=}$screening layer, layer 2$\rightarrow$ P${=}$plunger, S${=}$source, D${=}$drain gates, and layer 3$\rightarrow$ B${=}$Barrier gates) are 40 nm thick, while the Al$_2$O$_3$ oxide layers are 7 nm thick.}
    \label{fig1:device}
    \vspace{-0.6 cm}
\end{figure}
The modeled $z$-double well and DQD with relevant parameters are shown in Fig.~\ref{fig3a:dqdscheme}a. The second term in Eq.~\ref{eq1:spinvalleyHamiltonian} describes the spin-valley interaction:
\begin{eqnarray}\label{eq4:sigmatauHamiltonian}
&&H_{\sigma\tau}=\bigl(V(\mathbf{r})e^{-2ik_0z}\,\,\mathbb{I}_{2\times 2}\nonumber\\
&&+2\beta_0(k_+\sigma_++k_-\sigma_-)e^{2i(2\pi/a_{\text{Si}}-k_0)z}\bigr)\tau_-+h.c.,
\end{eqnarray}
where $\beta_0{=}8.2$ meV is the Dresselhaus SOI coefficient, calculated using many-body tight binding formalism~\cite{woods2023spin,woods2024g}. The third term in Eq.~\ref{eq1:spinvalleyHamiltonian} denotes the plunger gate potential, and the fourth term is the Zeeman splitting of the electron spin due to an external magnetic field $\mathbf{B}$. Here $\mathbf{g}$ denotes the $g$ tensor.
 \begin{figure}
    \centering
    \includegraphics[width=\linewidth]{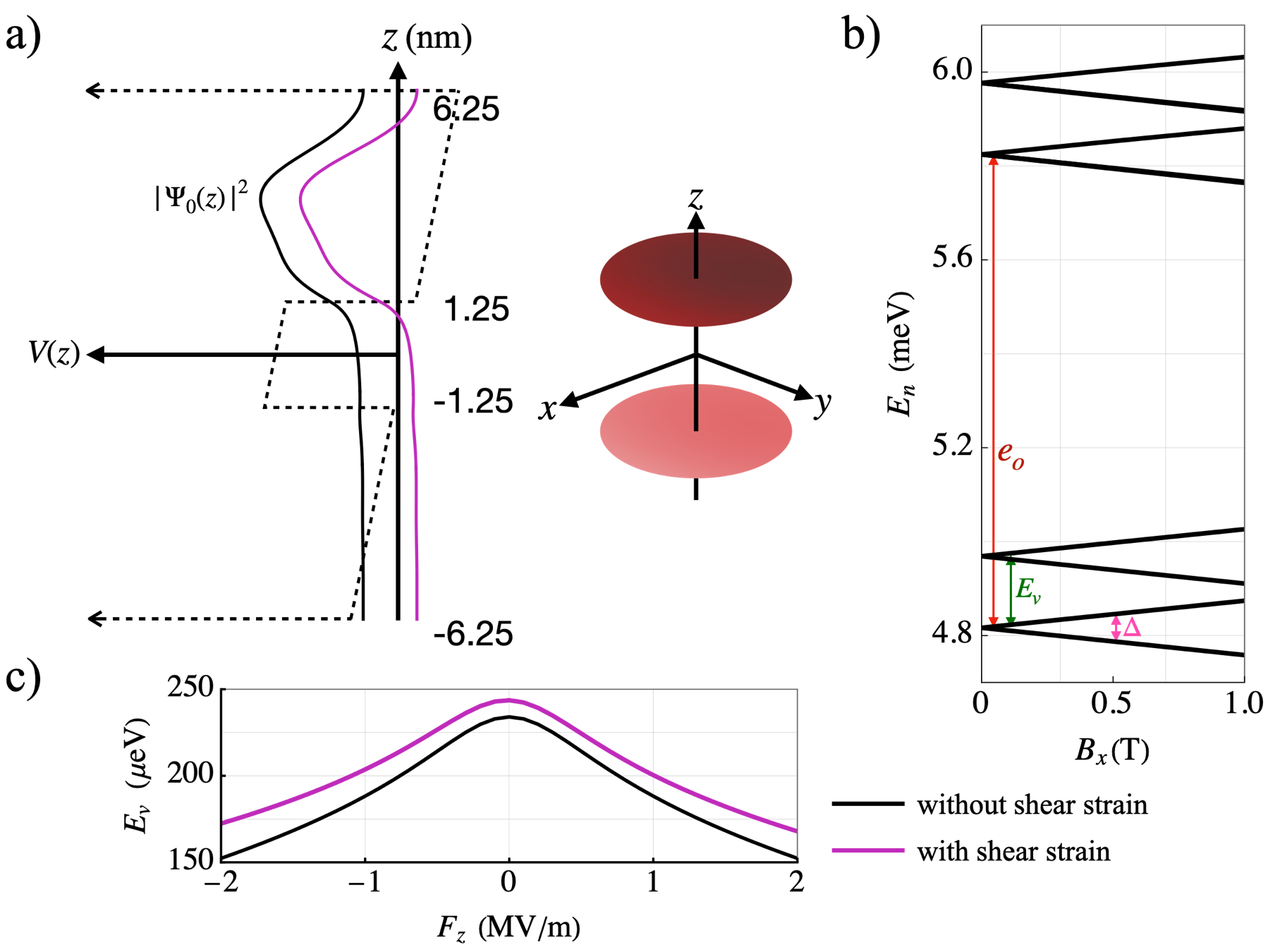}
    \vspace{-0.7 cm}
    \caption{{\bf The vertical DQD schematic.} a) Sketch of the vertical Si/Si$_{1-x}$Ge$_x$/Si DQD. The ground state $z$-wavefunction probability density $|\Psi_{0}(z)|^2$ at plunger field $F_z{=}-2$ MV/m, which resides mostly in the top Si well, with finite extent into the bottom well. Here, $L{=}5$ nm, $a{=}2.5$ nm, and $V_b{=}15$ meV ($x{=}0.033$). The tunnel coupling in the $z$-double well is $t_c{=}1$ meV. The in-plane confinement frequency is $\omega_{ip}{=}\hbar/\!\left(m_tL_{ip}^2\right)$, where $L_{ip}{=}20$ nm. b) The lowest spin, valley, and orbital states of the DQD without shear strain. Here, $\Delta{=}12\,\mu$eV, $E_v{=}250\,\mu$eV, and $e_0{=}1\,$meV denote the Zeeman, valley, and orbital energy splittings, {\it resp.} c) Valley splitting $E_v$ as a function of $F_z$ with (purple line) and without (black line) shear strain.
    }
    \label{fig3a:dqdscheme}
    \vspace{-0.6 cm}
\end{figure}
\begin{figure*}[ht!]
    \centering
    \includegraphics[width=\textwidth]{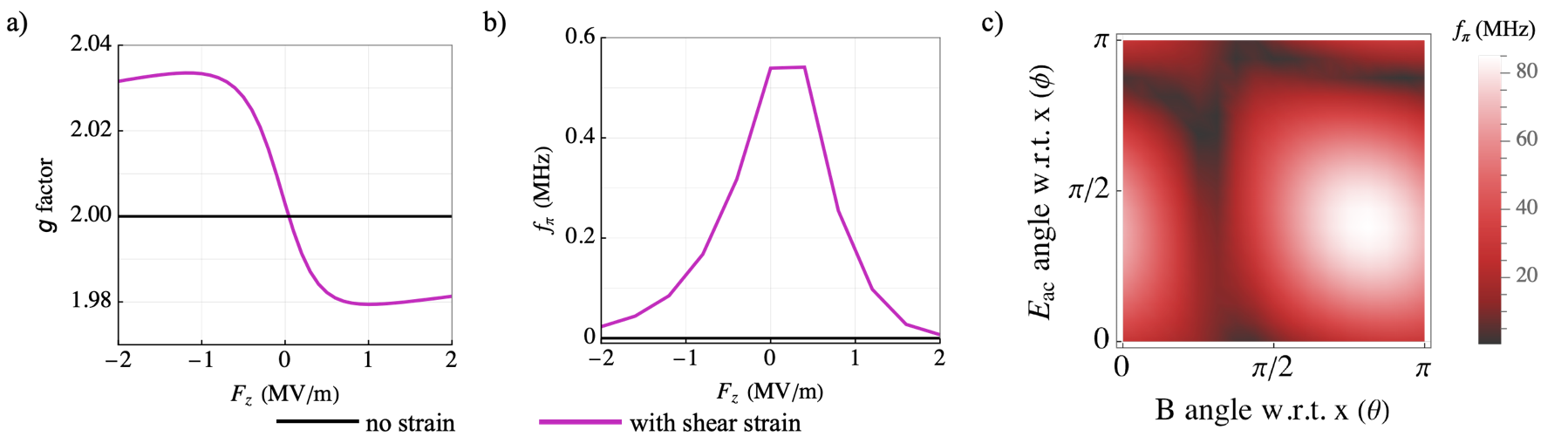}
    \vspace{-0.7 cm}
    \caption{\textbf{Electrical control of Si/Si$_{0.967}$Ge$_{0.033}$/Si DQD spin qubit.} a) The qubit $g$ factor as a function of the $P_1$ plunger field $F_z$. The variation around $g_0{=}2$ is a result of strong SOI in presence of the strain inhomogeneity. b) EDSR Rabi frequency $f_\pi$ vs. $F_z$ in presence of an ac electric field along $\mathbf{\hat{z}}$. In a) and b), specific scenarios without the shear strain (black solid line) and with shear strain (purple line) are shown. The magnetic field is $\mathbf{B}{=}(100,0,0)$ mT. c) 2D map of $f_\pi$ with the ac electric field in the $x{-}y$ plane and $F_z{=}-1$ MV/m. Here, $f_\pi$ is plotted as a function of the in-plane magnetic field angle $\theta$ ($\mathbf{B}{=}100(\cos\theta,\sin\theta,0)$ mT) and in-plane ac electric field angle $\phi$ ($\mathbf{E}_{\text{ac}}{=}E_{\text{ac}}^0 \cos(\omega t)(\cos\phi,\sin\phi,0)$ with $E_{\text{ac}}^0=10^4$ V/m).}. 
    \label{fig3:qubitelec}
    \vspace{-0.7 cm}
\end{figure*}

The eigenstates of the DQD spin qubit are given by $\left|\Psi_i\right\rangle=\sum_{lmnpq}c_i\,\phi_l(x)\,\phi_m(y)\,\phi_n(z)\left|\sigma_p\right\rangle\otimes\left|\tau_q\right\rangle$. The $z$-basis states $\phi_n(z)$ are the solutions to the 1D Schrödinger equation $\bigl[\hbar^2k_z^2/(2m_l){+}V_{\text{DW}}(z)\bigr]\phi_n(z){=}e_n\phi_n(z)$. Similarly, the $x{-}y$-basis states are given by harmonic oscillator states $\phi_l(x)\phi_m(y)$ \cite{suppl}. Choosing the basis sizes of $l{\in}\{1,2,3,4\}$, $m{\in}\{1,2,3,4\}$, and $n{\in}\{1,2,3,4,5,6,7,8\}$, the total Hamiltonian $H_{\text{tot}}\left|\Psi_i\right\rangle{=}E_i\left|\Psi_i\right\rangle$ of dimensions 512 by 512 can be diagonalized to evaluate the DQD qubit energy levels $E_i$ and the corresponding eigenfunction $\left|\Psi_i\right\rangle$. We denote the ground and first excited states as $\left|\Psi_0\right\rangle$ and $\left|\Psi_1\right\rangle$, {\it resp.} The contraction of the metal gatestack at cryogenic temperatures results in additional shear strain in the DQD, which modifies the SOI and enhances the valley splitting. To this end, we simulate the thermal contraction of a planar DQD device of Si/Si$_{1-x}$Ge$_x$ heterostructure with a tri-layer metallic gate stack on top (Fig.~\ref{fig1:device}). We employ the heat transfer module in COMSOL Multiphysics with relevant material parameters \cite{schaffler1997high,ioffe_SiGe_parameters} to generate maps of the inhomogeneous shear strain across the device (See SM Fig.~S1 \cite{suppl}). 

{\it Electrical tunability of the electron spin qubit.}
With full knowledge of the energy spectrum of a single electron spin in the vertical DQD (under $P_1$) with $L_{ip}{=}20$ nm, $L{=}5$ nm, $a{=}2.5$ nm, and $V_b{=}15$ meV ($x{=}0.033$), we now examine the spin qubit properties and their electrical tunability. The dependence of the lowest spin, valley, and orbital energy levels on an external magnetic field $\mathbf{B}{\parallel}\hat{\mathbf{x}}$ (Fig.~\ref{fig3a:dqdscheme}b) reveals a valley splitting of $E_v{\simeq}250\,\mu$eV and a Zeeman splitting of $\Delta{=}E_2{-}E_1{=}g\mu_B B{\simeq}10\,\mu$eV ($2.42$ GHz) for $|\mathbf{B}|{=}100$ mT. The large valley splitting originates from the vertical double well geometry, which enhances coherent coupling between the $+z$ and $-z$ valleys through interface induced Fourier components of the confinement potential near $2k_0$~\cite{stehouwer2025engineering,salamone2025valley,cvitkovich2025valley}. Random alloy disorder is not included in the effective mass approximation employed here, which captures the average band-edge physics of the Si/Si$_{1-x}$Ge$_x$ heterostructure. Atomistic studies have shown that alloy disorder can induce intervalley scattering and sample-dependent valley splittings~\cite{losert2023practical,thayil2025theory}. Nevertheless, in the vertical DQD, the valley splitting is primarily geometry-controlled rather than disorder-dominated. In fact, it can be shown that with realistic gate-induced strain in the vertical DQD, the valley splitting at finite plunger gate field is consistently large; exceeding $150\,\mu$eV across the relevant operating regime (see SM Fig.~S2~\cite{suppl}). Therefore, we expect alloy disorder to introduce device-to-device variability without qualitatively altering either the large valley splitting or the micromagnet-free electrical control mechanisms demonstrated in the proposed vertical DQD architecture.

As mentioned before, the lack of intrinsic SOI in Si implies that electrical manipulation of electron spin requires a nearby micromagnet; however, their large footprint and limited local tunability hinder the scalability of planar spin qubit devices. In the DQD device, strong SOI originates from (i) the atomistic Dresselhaus-type spin-valley interaction, as given by the second term in Eq.~\ref{eq4:sigmatauHamiltonian}; (ii) the spatially varying gate-induced strain, which effectively produces an inhomogeneous electric field that the dot wave function experiences when shaken; and (iii) the structural inversion asymmetry (SIA) introduced by the plunger gate field $F_z$. We demonstrate that the SOI in the vertical DQD opens up multiple micromagnet-free approaches to electrical spin manipulation. Firstly, the presence of shear strain changes the $g$ factor by $\delta g\approx 1\%$ when the field of the top/plunger gate is varied from negative to positive, essentially pushing the qubit wavefunction into the bottom or the top Si well (Fig.~\ref{fig3:qubitelec}a). Although small, this $g$ factor tunability via the plunger gate allows one to turn the spin rotations on and off electrostatically. The $\delta g{\approx}1\%$ change corresponds to the $10$ MHz shift in qubit Larmor frequency, which can help avoid frequency-dependent impedance mismatches and standing wave resonances in the microwave transmission line, often separated by $\approx10$ MHz. Additionally, this tunability implies that local control in quantum dot arrays would be possible exclusively via the plunger gates. Moreover, for $\mathbf{B}{=}B_x\hat{\mathbf{x}}$, an out-of-plane ac electric field $\mathbf{E}_{\text{ac}}{=}E^0_{\text{ac}}\cos(\omega t)\mathbf{\hat{z}}$ supplied through the plunger gate can induce a transverse magnetic field ${\propto}\,B_x\frac{\partial g_{xy}}{\partial F_z}E_{\text{ac}}^0\cos(\omega t)\hat{z}\sigma_y$. This leads to $g$-tensor modulation resonance ($g$-TMR), a particular mechanism of EDSR where the spin exhibits Rabi oscillations when driven in a changing $g$ tensor. We evaluate $f_\pi{=}\,550$ kHz for $\mathbf{E}_{\text{ac}}{\parallel}\mathbf{\hat{z}}$ (Fig.~\ref{fig3:qubitelec}b), where the EDSR Rabi frequency $f_\pi$ in the presence of an ac electric field $\mathbf{E}_{\text{ac}}{=}E^0_{\text{ac}}\cos(\omega t)\mathbf{\hat{n}}$ is calculated as follows:
\begin{eqnarray}
hf_\pi=eE^0_{\text{ac}}\left|\left\langle\Psi_0\right|\mathbf{\hat{n}}\cdot\mathbf{r}\left|\Psi_1\right\rangle\right|,
\end{eqnarray}
where $E^0_{\text{ac}}$ is the amplitude and $\mathbf{\hat{n}}$ is the direction of the ac drive. Note that $f_\pi$ for $\mathbf{E}_{\text{ac}}{\parallel}\mathbf{\hat{z}}$ can be enhanced by improving the $g$-factor gradient through geometric means, up to a few MHz (see SM Fig.~S3~\cite{suppl}). The  amplitude is assumed to be $E_{ac}^0{=}10^4$ V/m~\cite{nowack2007coherent,kawakami2014electrical}. 

Notably, ESR due to a fixed frequency ac magnetic field $\mathbf{B}_{\text{ac}}{=}B_\text{ac}^0\cos(\omega_\text{\tiny{ESR}}t)$ applied perpendicular to $\bf B$ can be electrically controlled via local gates by exploiting the $g$-factor variation in the vertical DQD. Indeed, by shifting the electron wave function, the Zeeman splitting $\Delta$ of its spin can be tuned to resonance so that $\Delta=\hbar\omega_{\text{\tiny{ESR}}}$. Typical ESR powers of $5$ dBm in experiments yield an ac magnetic field amplitude of $0.1$ mT after device dissipation, leading to $f_\pi{=}2.8$ MHz~\cite{chan2018assessment}. Achieving resonance control requires tuning $\Delta$ beyond the ESR linewidth (${\sim}\,100$ KHz~\cite{steinacker2025industry}), which is 
satisfied for $\delta g {\sim}\, 0.01$ and typical $\omega_\text{\tiny{ESR}}{\sim}\,5$ GHz.
An in-plane ac electric field $\mathbf{E}_{\text{ac}}{=}E^0_{\text{ac}}\cos(\omega t)(\cos\phi\,\hat{x}+\sin\phi\,\hat{y})$ could facilitate strong iso-Zeeman EDSR. Here, the qubit is driven at a constant $g$ factor via a transverse magnetic field originating from the net effect of SOI and $\mathbf{E}_{\text{ac}}$. We calculate fast EDSR of up to $f_\pi{=}80$ MHz with an in-plane ac electric field (Fig.~\ref{fig3:qubitelec}c). While iso-Zeeman EDSR is significantly faster than $g$-TMR, the latter is more robust against charge noise. This is due to the reliance of iso-Zeeman EDSR on SOI, which exposes the qubit to electric-field fluctuations from nearby charge traps. Therefore, $g$-TMR at higher magnetic fields could potentially lead to full electrical control of the vertical DQD spin qubit using only the plunger gate~\cite{ivlev2025operating}. 
\begin{figure}[htbp!]
\centering
\includegraphics[width=\linewidth]{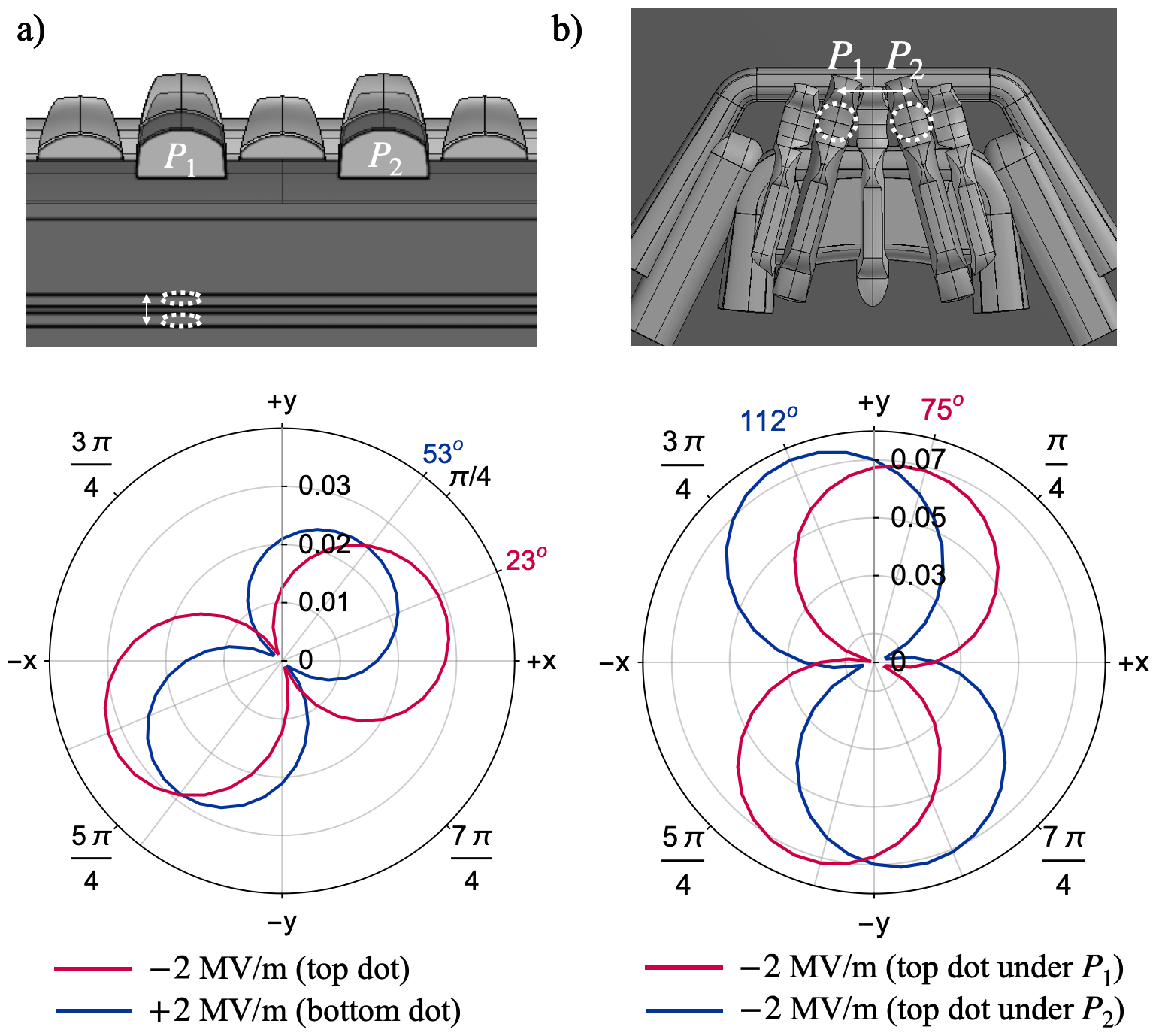}
\vspace{-0.6 cm}
\caption{\textbf{$g$-factor anisotropy in DQDs in the presence of gate-induced strain.} The quantity plotted is $|g(\theta)-2|$, where $g(\theta){=}\frac{\Delta(\theta)}{\mu_BB}$, and the Zeeman splitting $\Delta$ varies with the magnetic field angle, $\mathbf{B}{=}B(\cos\theta,\sin\theta,0)$. a) The polar plot of $|g(\theta)-2|$ for a vertical DQD under plunger gate $P_1$ with $F_z{=}2$ MV/m and $F_z{=}-2$ MV/m. b) Polar plot of $|g(\theta)-2|$ for left (under $P_1$) and right (under $P_2$) dots with rotated plunger gates at $F_z{=}-2$ MV/m. The plunger gates $P_1$ and $P_2$ are rotated by $30^\circ$ with respect to each other.} 
\label{fig4:spinhopping}
\vspace{-0.4 cm}
\end{figure}

Lastly, we examine the in-plane $g$-factor anisotropy, which can facilitate single qubit rotations via spin hopping~\cite{loss1998quantum,wang2024operating,unseld2025baseband}. Including gate-induced strain, we present the polar plot of the deviation of the $g$ factor from $g_0{=}2$ for a sweep of $B_x$-$B_y$, where $|\mathbf{B}|{=}100$ mT. The directions of maximum and minimum $|g(\theta){-}2|$ define the principal axes of the in-plane $g$ tensor. A pulse can be engineered to periodically shuttle the qubit between the top and bottom dots with wait times $\tau_b$ at the bottom dot and $\tau_t$ at the top dot (see SM Fig.~S4~\cite{suppl}). We calculate the difference between $g$-tensor principal axes to be ${\sim}\,30^{\circ}$, so a $\pi/2$ spin rotation is achieved after 2 $(\tau_t{+}\tau_b)$ sequences consisting of 3 hoppings, with an $S$-gate time of $T_g{=}\,0.6$ ns. Shuttling in the $x$-direction, between the top dots under $P_1$ and $P_2$ plunger gates, is considered as well. We find that when the plunger gates are tilted  by $30^{\circ}$ with respect to each other, a $\,37^{\circ}$ difference is produced in the in-plane principal  axes of the $g$ tensor. The deviation of the $g$ factor from $g_0{=2}$ is also enhanced. Horizontal DQD shuttling has the advantage of requiring smaller pulse voltages \cite{wang2024operating}. In contrast, vertical DQD shuttling requires larger pulse amplitudes but offers the benefit of ultrafast qubit manipulation using a single plunger gate.

{\it Conclusion.}
We have shown that vertical DQDs in Si/Si$_{1-x}$Ge$_x$ heterostructures enable large valley splitting and exhibit micromagnet-free full electrical control of an electron spin qubit. An in-plane ac electric field generates fast EDSR with a gate time of ${\sim}30$ ns. An out-of-plane ac electric field supplied via the top-gate also produces $100$ KHz Rabi frequency. Combined with the $g$-factor tunability, this potentially provides a mechanism by which the spin qubit can be entirely controlled via the plunger gate. We also establish that the angular difference in the $g$-tensor principal axes between vertically as well as horizontally neighboring quantum dots facilitates spin-hopping-based spin rotation. The path forward is to study the two-spin problem and investigate the feasibility of universal gate operations in this architecture.

{\it Acknowledgments}--
We thank Ji Zou, Zolt{\'a}n Gy{\"o}rgy, and Stefano Bosco for fruitful discussions and useful comments. This work was supported in part by NCCR SPIN, a National Center of Competence in Research funded by the Swiss National Science Foundation (grant number 225153). 
This publication is based on work supported by King Fahd University of Petroleum \& Minerals. The author at KFUPM acknowledges the Deanship of Research and the Center for Advanced Quantum Computing for the support received under Grant no. CUP25102 and no. INQC2500, respectively.

{\it Data availability}-- The data that support the findings of this article are openly available \cite{sarkarmicrodata2025}.
\bibliographystyle{apsrev4-1}
\bibliography{SiDQDref}
\end{document}